\documentclass[useAMS,usenatbib]{mn2e}

\usepackage{graphicx}
\usepackage{natbib}

\newcommand{\km}{\,\mbox{km}\,\mbox{s}^{-1}}

\def\Ha{H$\alpha$}

\newcommand{\mnras}{MNRAS}

\title[What controls the ionized gas turbulent motions in dwarf galaxies?]{What controls the ionized gas turbulent motions in dwarf
  galaxies?}  \author[Moiseev et al.]
{Alexei V. Moiseev$^1$\thanks{moisav@gmail.com}, Anton V. Tikhonov$^2$\thanks{Deceased},  and Anatoly Klypin$^3$\\
  $^1$Special Astrophysical Observatory, Russian Academy of Sciences\thanks{The system of Russian Academy of Sciences institutes was liquidated on Sep 2013},  369167 Nizhnii Arkhyz,  Karachaevo-Cherkesskaya Republic,  Russia\\
  $^2$St. Petersburg State University, Universitetskii pr. 28,  198504 St. Petersburg, Stary Peterhof,  Russia\\
  $^3$New Mexico State University, USA }

\begin{document}

\date{Accepted ....  Received ....}

\pagerange{\pageref{firstpage}--\pageref{lastpage}} \pubyear{2011}

\maketitle

\label{firstpage}

\begin{abstract}
Using 3D spectroscopy with a scanning Fabry--Perot
interferometer, we study the ionized gas kinematics in 59 nearby
dwarf galaxies. Combining our results with data from literature, we
provide a global relation between the gas velocity dispersion
$\sigma$ and the star formation rate (SFR) and \Ha\, luminosity
for galaxies in a very broad range of star formation rates
SFR=$0.001-300\,M_\odot\,{\rm yr}^{-1}$. We find that the
$SFR - \sigma$ relation for the combined sample of dwarf galaxies,
star forming, local luminous, and ultra-luminous infrared galaxies
can be fitted as $\sigma\propto SFR^{5.3\pm0.2}$. This implies that the
slope of the $L-\sigma$ relation inferred from the sample of rotation
supported disc galaxies (including mergers) is similar to the $L-\sigma$ relation
of individual giant HII regions. We present
arguments that the velocity dispersion of the ionized gas does not
reflect the virial motions in the gravitational potential of dwarf
galaxies, and instead  is mainly
determined by the energy injected into the interstellar medium by
the ongoing star formation.
\end{abstract}

\begin{keywords}
galaxies: dwarf -- galaxies: kinematics and dynamics --
galaxies: ISM -- ISM: bubbles.
\end{keywords}

\maketitle

\section{Introduction}

The nature of high-velocity turbulent motions of ionized gas in giant
star-forming regions and dwarf galaxies has been studied for almost
half a century, starting with \citet{SmithWeedman1970}, who showed
that the velocity dispersion, $\sigma$ (the rms velocity along
line-of-sight)  in giant star forming regions of the M33 and M101
galaxies, determined from the width of the integral \Ha\ emission line
profiles is about a few tens of $\km$. A close correlation was later
found between the rms velocity $\sigma$, the diameter, and the total
Balmer line luminosity (in the $\mbox{H}\beta$ emission line) of the
emitting nebula \citep{Melnick1977, TerlevichMelnick1981}.  Similar
trends were found for individual HII regions and for dwarf irregular
(dIrr) and blue compact (BCDG or HII) galaxies. Because of the
tightness of the observed $L(\mbox{H}\beta)-\sigma$ relation of
star-bursting compact galaxies, it was proposed as an independent
indicator of the cosmological distance
\citep[e.g.][]{Melnick1987,Melnick2000,Chavez2012,Koulouridis2013,Chavez2014}.

Relation between the emission-line luminosity $L$ and ionized gas
 $\sigma$ is also important for understanding how star
formation affects motion of gas and how star formation is regulated by
different stellar feedback processes.
Although the existence of a close correlation between these quantities
has been known for a long time, the origin of luminosity--velocity dispersion
relation in HII galaxies and giant HII regions remains unclear
\citep{TerlevichMelnick1981,ChuKennicutt1994,Scalo1999,Bordalo2009,MoisLoz2012}.
The following factors may affect the width of the observed ionized hydrogen lines:

\begin{enumerate}
\item Thermal broadening, which amounts to $\sigma_{th}\approx8-10\km$
  for typical electron temperatures in the HII regions 7\,000--10\,000 K.

\item Turbulent motions determined by the combined influence of young
  stellar clusters on the interstellar medium (ISM).

\item Gravitational broadening, caused by virial motions of gas clouds
  in galaxy gravitational potential.

\item  Non-virial gravitational motions: ISM turbulence
    associated with tidal interactions, galaxy merging and external
    gas accretion (e.g. numerical simulations by \cite{Bournaud2011},
    observational constrains in \cite{Arribas2014} and references
    therein.)    

\end{enumerate}

When presenting and discussing the $L-\sigma$ relation, we
will always assume, except in specially mentioned cases, that $L$ is
the total  luminosity of a galaxy or a HII region in the \Ha\ line, and
$\sigma$ is the average (luminosity-weighed) velocity dispersion of ionized gas.

It is often assumed that the gravitational effects are the dominante
factor in giant HII regions and HII galaxies
\citep[e.g.,][]{TerlevichMelnick1981,Tenorio-Tagle1993,Melnick1987}.
The main argument was that the  velocity dispersion of ionized gas
is largely determined by the wind of stars that participate in
virial motions with the characteristic velocity
$\sigma_{\rm stars}$. Then $\sigma$ and $\sigma_{\rm stars}$ are
mainly controlled by the gravitational potential of the
object, and the $L-\sigma$ relation is similar to the
Faber-Jackson relation for elliptical galaxies:
$L\propto\sigma_{\rm stars}^4$. Different studies of the ionized gas
kinematics in giant extragalactic HII regions give a value between
$\sim3$ and $\sim7$ for the exponent in the $L-\sigma$ relation
\citep{Blasco-Herrera2010,Blasco-Herrera2013}.
\citet{Chavez2014} present a detailed analysis of this
relation for a sample of 128 local compact HII galaxies. They
demonstrate that adding the second (the size of HII regions) or even
the third (the emission line equivalent width  $EW$ or the continuum
color and metallicity) parameters significantly improve the
correlation. They also argue in favor of the gravitational origin
of the ionized gas velocity dispersion $\sigma$. They also emphasize that in order the $L-\sigma$ to be tight,
the regions of recent  bursts of star
formation  should be gravitationally bound, compact, massive, and
have strong emission lines with $EW(H\beta)>50$\AA with pure Guassian proile
without any evidences of multiplicity or rotation.

A different view on the $L-\sigma$ relation was developed beginning
with \citet{GallagherHunter1983}, who suggested that the processes
related with the energy of embedded OB stars drive the ionized gas
velocity dispersion in giant HII regions on scales smaller than
0.5~kpc, while properties of larger supergiant HII complexes agree
with a gravitationally driven $\sigma$. Based on Fabry--Perot interferometric
observations of nearby galaxies \citet{Arsenault1988} also concluded
that effects of stellar wind and turbulence are more important for
the kinematics of giant HII regions compared with virial
motions. There are different ways of how a young stellar population
affects the surrounding ISM. According \citet{MacLowKlessen2004},
\citep[see also][and references therein]{Lopez2014} the main
mechanisms are: protostellar outflows, stellar winds and ionizing
radiation pressure of massive stars, supernovae (SN) explosions, the
dust-processed infrared radiation, and warm and hot gas
pressure. The contribution of different factors changes with spatial
and density scale. For instance, \citet{Lopez2014} found that warm
ionized gas dominates over the other terms of pressure in all
considered HII regions in Large and Small Magellanic
Clouds. Numerical simulations also demonstrate that the radiation
pressure is a very important feedback mechanism for models of
formation of galaxies compared with the hot gas contribution (heated
by SNs and stellar winds)
\citep[e.g.,][]{Hopkins2012,Ceverino2014,Trujillo2015}.  Note that
together with the chaotic gas motions, responsible for the Gaussian
emission line profile, the effects related with individual expanding
shells may lead to the appearance of non-symmetric features such as
wings and peaks in the line profile \citep*[see examples
in][]{Melnick1999,BordaloTelles2011}.

Some recent observations also provide evidence that  energy
injected into the interstellar medium by the ongoing star formation
process is the main factor affecting  gas turbulent
motions. For example, \citet{Green2010} show that in a wide range of
galaxy luminosities the ionized gas rms velocity $\sigma$ is determined by the star
formation rate (SFR), which is proportional to the \Ha~ luminosity, and
does not correlate with the galaxy mass. Earlier, \citet*{Dib2006}
 showed that $\sigma$ for neutral gas also depends on the SFR.
However other  numerical simulations with higher
resolution in a stratified ISM suggested that this trend is absent
if the gas surface density increases with the SN rate
\citep{Joung2009}. On the other hand, the galaxy-scale simulations  that included stellar
feedback  \citep{Hopkins2012} clearly demonstrate
that  the average velocity dispersion of the gas (in all
cold, warm and hot phases) increases with the total SFR.  Also,
\citet{Dopita2008} argues that kinetic energy of ionized gas in the
star formation regions is proportional to the local SFR, integrated
over the duration of the burst.  \citet{MoisLoz2012} demonstrate a
close correlation between the two-dimensional distribution of the
radial velocity dispersion of ionized gas and the surface brightness
in the \Ha\ line for a sample of nearby dwarf galaxies:  most of
the regions with the highest velocity dispersion belong to a
low-brightness diffuse background surrounding  large
HII-regions.Recent simulations of multiple SN explosions were able
to reproduce the diagrams `\Ha{} intensity -- velocity dispersion'
observed in these galaxies when a realistic spatial resolution was
 used \citep*{Vasiliev2015}.

However, some other recent studies of high-redshift galaxies
contradict \citet{Green2010}. For instance,
integral-field data by \citet{Genzel2011} demonstrate a very weak
correlation of $\sigma$ and SFR density for star-forming clumps
in galaxies at $z=2.2-2.4$. Further \citet{Wisnioski2012} and
\citet{Swinbank2012} using observations with similar technique
argue that such clumps follow the same $L-\sigma$ relation as
gravitational bounded local giant HII regions according
\citet{TerlevichMelnick1981} and related studies.

To summarize, the question of the nature of high-velocity turbulent
motions of ionized gas in dwarf galaxies still remains open. One of
the problems is the lack of sufficiently uniform observational
data. Recent results by \citet{BordaloTelles2011} fill this gap to a
certain extent. They give a uniform set of high-resolution
spectroscopy observations for 118 star formation regions in HII
galaxies. \citet{Chavez2014} presented  similar data for  128
HII galaxies selected from the SDSS. They also discussed the impact of other
factors (gas metallicity, ionization state, the history of star
formation) on the $L-\sigma$ relation.

However, most of the results are based on slit spectroscopy.  The
distribution of ionized gas in  dwarf galaxies has a complex
irregular morphology, and the two-dimensional velocity dispersion
maps, obtained using the 3D spectroscopy can provide the most complete
information about gas turbulence.

The second important issue is related with the fact that the
luminosity-weighted velocity dispersion has become a widely used
value to characterize the turbulent gas motions {in  samples of   distant and nearby galaxies \citep{Ostlin2001,Green2010,Davies2011,Blasco-Herrera2013,Arribas2014}.} How does the
$L-\sigma$ relation work for galaxies of different types and
luminosities? Do galaxies obey the same scaling
relation as the HII regions and star forming clumps?

In this paper we analyse ionized gas turbulent motions in 59 galaxies observed at the
6-m Big Telescope Alt-azimuthal (BTA) of Special Astrophysical
Observatory of the Russian Academy of Sciences (SAO RAS) using a
scanning Fabry--Perot interferometer (FPI). Most of our sample consists
of dwarf galaxies of the Local Universe. Using this unique material, we
were able to extend the $L-\sigma$ relation for objects much weaker
than in the samples described earlier.

\begin{figure*}
\centerline{
\includegraphics[width=0.32\textwidth]{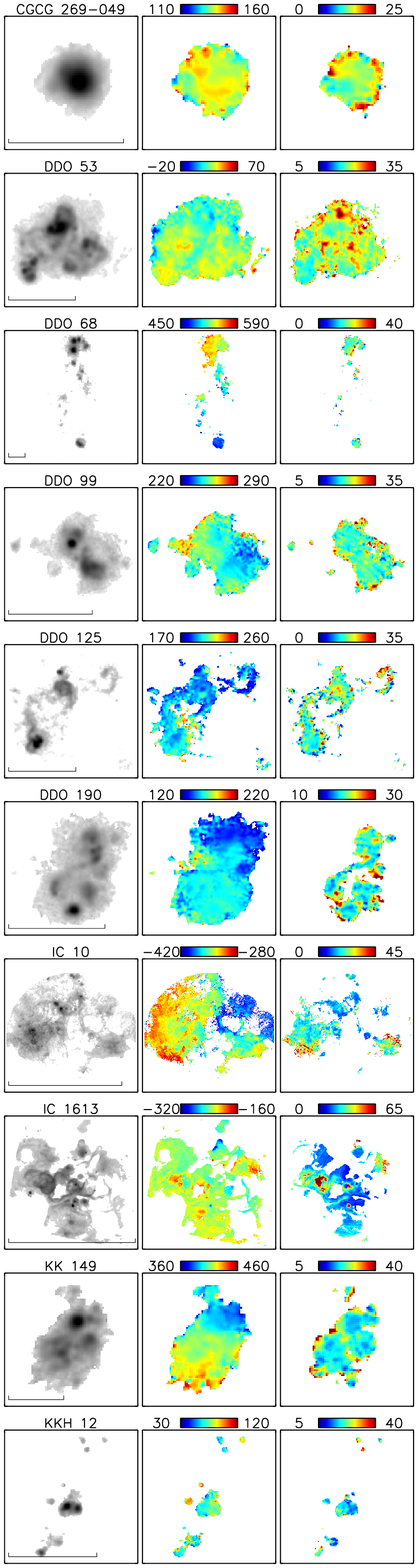}
~
\includegraphics[width=0.32\textwidth]{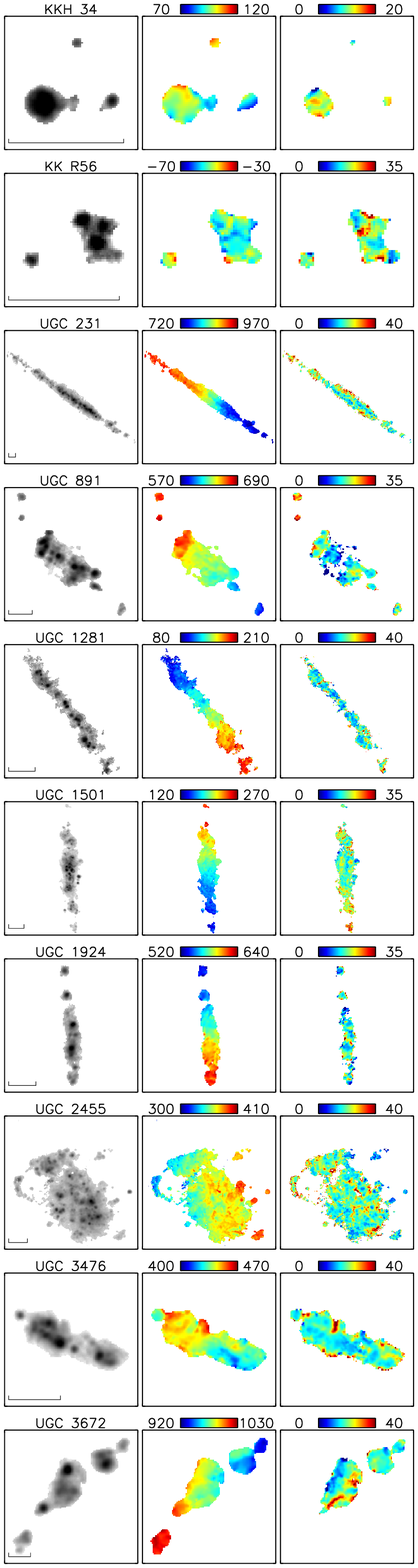}
~
\includegraphics[width=0.32\textwidth]{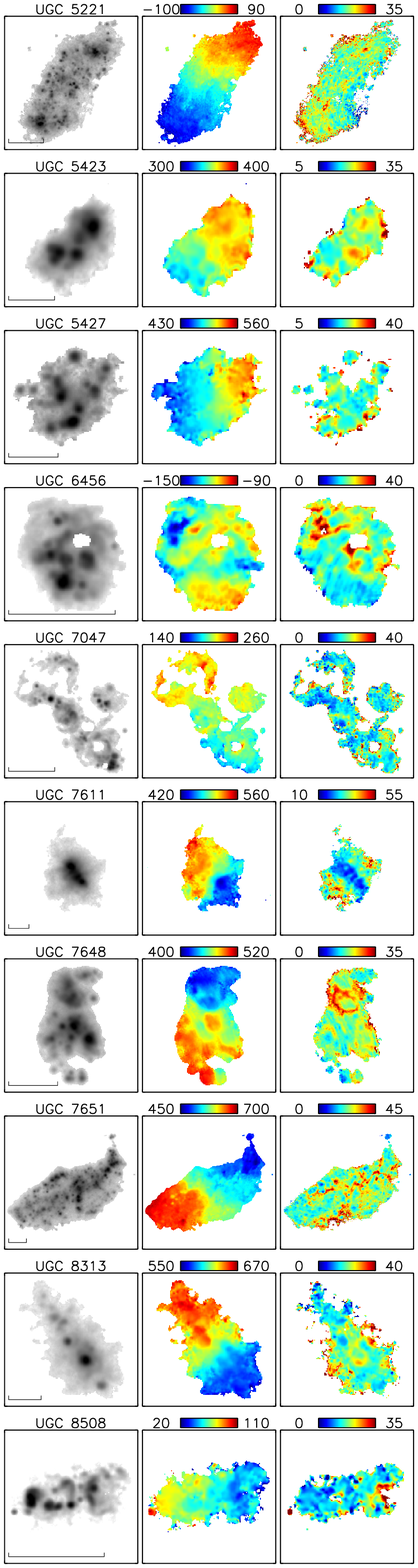}
} \caption{Results of observations with the scanning FPI at the SAO
  RAS 6-m telescope. Given for each galaxy: an image in the \Ha{}
  line, maps of line-of-sight velocities, and velocity dispersion,
  corrected for the thermal line broadening. Colour scale is in $\km$. The
  horizontal bar in the first panel for each galaxy shows the linear scale of 1 kpc.}
\label{fig1_1}
\end{figure*}

\setcounter{figure}{0}
\begin{figure*}
\centerline{
\includegraphics[width=0.32\textwidth]{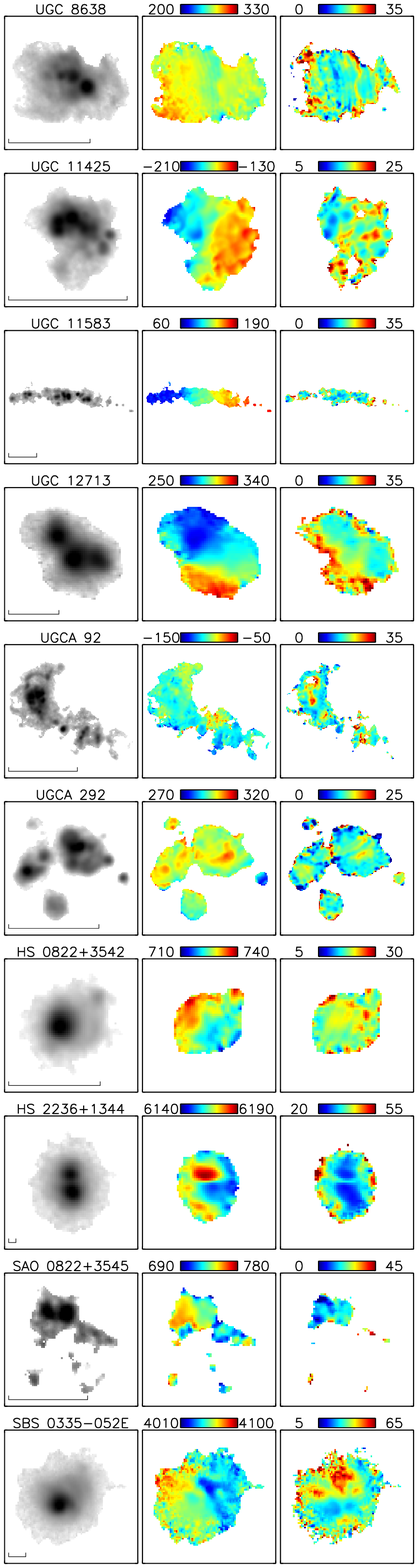}
~
\includegraphics[width=0.32\textwidth]{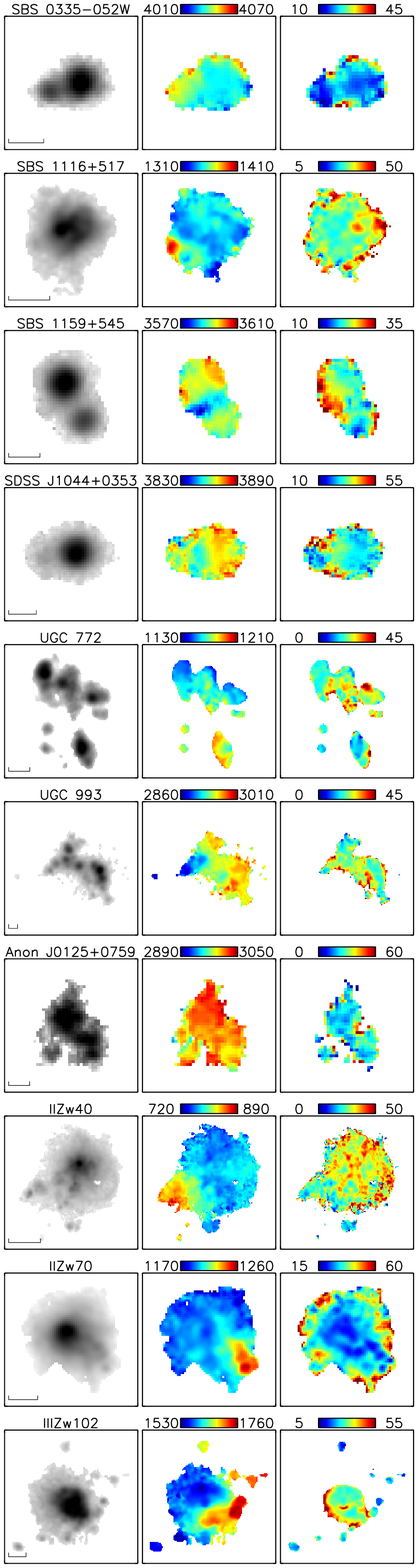}
~
\includegraphics[width=0.32\textwidth]{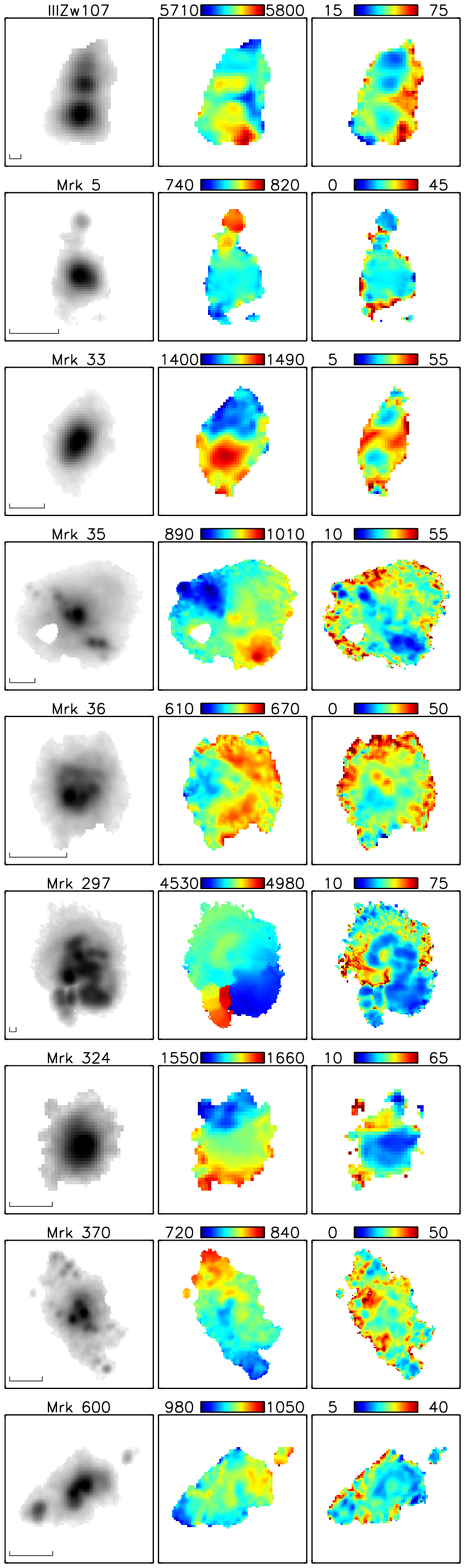}
}
 \caption{(continue)}
\end{figure*}

\section{Observations and data reduction}
\label{obs}

\subsection{Sample of galaxies}

Our sample   is based on the archive data of observations at the
6-m telescope of SAO RAS. It consists of several subsamples of
galaxies studied within the framework of several observational
programs:

\begin{itemize}

\item The Local Volume (LV) subsample: 36 nearby dwarf galaxies
  with galactocentric distances $D<15$ Mpc. These are mostly
  low-luminosity objects with the average absolute magnitude
  $M_B\approx-14$.  The  ionized gas velocity   dispersion maps of 10  galaxies from this subsample
  were already reported  by \citet{MoisLoz2012}.

\item The XMD subsample of 9 more distant ($D=14-86$ Mpc) and  bright
  ($M_B\approx-15$) dwarf galaxies with extremely low metallicity, and
  2 low surface brightness companion galaxies. Their internal
  kinematics were described and analysed in \citet*{Moiseev2010}.

\item The BCDG subsample: 12 blue compact dwarf galaxies at distances $D=10-78$
  Mpc; all are brighter than $M_B=-15$. Some of them (e.g., Mrk 297 or
  III~Zw~102) are sometimes included in BCDG samples
  \citep[see][]{Cairos2001}, though they cannot be formally called
  dwarf galaxies since their absolute magnitude is brighter than
  $M_B=-20$ and the amplitude of the rotation curve exceeds
  $140\km$. The observational data for 3 of these galaxies were
  presented by \citet{delgado}.
  \end{itemize}

The full sample consists of 59 galaxies, and covers a wide
range of luminosities from $M_B=-11$ to $M_B=-21$. The overwhelming majority
(51/59=86 per cents) of them are dwarf galaxies with the absolute
magnitudes  $M_B>-18$.

The observations were carried out in the prime focus of the 6-m
telescope of SAO RAS using a scanning FPI mounted inside the SCORPIO
focal reducer \citep{AfanasievMoiseev2005}. In 2014 its  new version  SCORPIO-2 \citep{AfanasievMoiseev2011} was used. The operating spectral
range around the H$\alpha$ line was cut using a narrow bandpass
filter. About $\sim2/3$ of the observations were performed using the FPI501
interferometer, providing in the H$\alpha$ emission line a free
spectral range between the neighboring interference orders
$\Delta\lambda=13$\,\AA\, and spectral resolution ($FWHM$ of the
instrumental profile) of about $0.8$\AA\, ($35\km$), sampled by
0.36\AA\, per channel. After November 2009 we used a new FPI751
interferometer, which has $\Delta\lambda=8.7$\,\AA\, and a spectral
resolution of $0.4$\AA\, ($18\km$), sampled by 0.21\AA\ per
channel. The Mrk 33 galaxy was observed in the [NII]$\lambda6583$
line, all the others -- in the \Ha\ line.

In 2003-2014, the detectors were CCD EEV 42-40 and E2V 42-90,
given the image scale of $0.71\,\mbox{arcsec}\, \mbox{pixel}^{-1}$ in
the on-chip binned $4\times4$ mode. In 2002 the TK1024 CCD was used,
yielding the  scale of $0.56\,\mbox{arcsec}\,\mbox{pixel}^{-1}$
in the $2\times2$ binning mode.

During the scanning process we have consistently obtained 36
interferograms (40 for FPI751) at different distances between the FPI
plates. The seeing on different nights varied from 1 to 4 arcsec. The
reduction of observational data was performed using the software
package running in the IDL environment
\citep{Moiseev02ifp,MoiseevEgorov2008}. Following the primary
reduction, airglow lines subtraction, photometric and seeing
corrections using the reference stars and wavelength calibration, the
observational data were combined into the data cubes, where each pixel
in the field of view contains a 36- or 40-channel spectrum.

The log of observations is shown in Table~\ref{tab_obs}, listing the
galaxy name; the date of observations; the
interferometer type; exposure time; the resulting angular resolution
after smoothing the reduced data cubes with a two-dimensional Gaussian to
increase the signal-to-noise ratio in the areas of low surface
brightness. Only the information for the Local Volume galaxies and the BCDG
subsamples is shown; see the log of observations in \citet{Moiseev2010} for the XMD subsample.

\begin{table*}
\caption{Log of observations}\label{tab_obs}
\begin{tabular}{lcccc}
\hline
Name         & Date     & FPI    & Exp. time    & Ang.resol. \\
             &          &        &  time (s)    &  (arcsec)  \\
\hline
\multicolumn{5}{c}{Local volume subsample}\\
CGCG269-049& 06.02.2010 & FPI751 & $150\times40$& 4.4  \\
DDO~53     & 26.02.2009 & FPI501 & $200\times36$&  3.3   \\
DDO~68     & 30.12.2006 & FPI501 & $240\times36$&  2.7   \\
DDO~99     & 26.02.2009 & FPI501 & $180\times36$&  3.8   \\
DDO~125    & 18.05.2005 & FPI501 & $180\times36$&  3.0   \\
DDO~190    & 04.03.2009 & FPI501 & $100\times36$&  3.3   \\
IC~10      & 08.09.2005 & FPI501 & $300\times36$&  1.2   \\
IC~1613    & 12.09.2001 & FPI501 & $200\times36$&  2.2   \\
KK~149     & 05.03.2009 & FPI501 & $150\times36$& 2.8   \\ 
KKH~12     & 23.08.2004 & FPI501 & $120\times36$&  2.7   \\
    KKH~34 & 12.11.2009 & FPI751 & $230\times40$ & 3.4   \\
KKR~56     & 20.05.2010 & FPI751 & $150\times40$ & 3.3   \\
  UGC~231  & 11.11.2009 & FPI751 & $200\times40$ & 2.4   \\
 UGC~891   & 11.11.2009 & FPI751 & $200\times40$ & 3.1   \\
UGC~1281   & 14.08.2009 & FPI751 & $110\times40$ & 2.7   \\
UGC~1501   & 10.11.2009 & FPI751 & $200\times40$ & 2.3   \\
UGC~1924   & 11.11.2009 & FPI751 & $180\times40$ & 2.3   \\
 UGC~3476  & 02.11.2010 & FPI751 & $220\times40$ & 2.4   \\
 UGC~3672  & 12.11.2009 & FPI751 & $160\times40$ & 3.1   \\
 UGC~5221  & 16.12.2014 & FPI751 & $160\times40$ & 2.0   \\
 UGC~5423  & 26.02.2009 & FPI501 & $180\times36$ & 3.5   \\
 UGC~5427  & 04.03.2009 & FPI501 & $180\times36$ & 3.7   \\
 UGC~8638  & 24.02.2009 & FPI501 & $150\times36$ & 3.9   \\
UGC~6456   & 29.11.2002 & FPI501 & $300\times36$&  2.2   \\
UGC~7611   & 19.05.2010 & FPI751 & $ 160\times40$ &3.5    \\
UGC~8508   & 16.05.2005 & FPI501 & $200\times36$&  3.0   \\
UGC~11425  & 14.08.2009 & FPI751 & $140\times40$ & 3.0   \\
UGC~11583  & 10.11.2009 & FPI751 & $220\times40$ & 1.9   \\
UGC~12713  & 16.05.2005 & FPI501 & $200\times36$&  3.0   \\
UGC~2455   & 07.10.2010 & FPI751 & $240\times40$&  2.3   \\
UGC~7047   & 17-18.03.2012 & FPI751 & $360\times40$&  2.2   \\
UGC~7648/51   & 08.02.2011 & FPI751 & $160\times40$&  2.8   \\
UGC~8313   & 19.03.2012 & FPI751 & $450\times40$&  2.7   \\
UGCA~92    & 10.11.2009 & FPI751 & $180\times40$&  2.5   \\
 UGCA~292  & 07.02.2010 & FPI751 & $120\times40$ & 3.6   \\
\multicolumn{5}{c}{BCDG subsample}\\
   II~Zw~40 & 01.12.2003 & FPI501 & $120\times36$&  1.9   \\
   II~Zw~70 & 30.01.2004 & FPI501 & $180\times36$&   2.5  \\
 III~Zw~107 & 01.12.2006 & FPI501 & $100\times36$&  1.9   \\
 III~Zw~102 & 30.11.2003 & FPI501 & $180\times36$&  2.7   \\
      Mrk~5 & 30.01.2004 & FPI501 & $240\times36$&   2.6     \\
     Mrk~33 & 30.01.2004 & FPI501 & $150\times36$&   2.3     \\
     Mrk~35 & 01.12.2003 & FPI501 & $150\times36$&   2.1     \\
     Mrk~36 & 29.11.2003 & FPI501 & $160\times36$&   1.6     \\
    Mrk~297 & 13.08.2009 & FPI751 & $144\times40$&  3.5      \\
    Mrk~324 & 30.11.2003 & FPI501 & $120\times36$&   2.8     \\
    Mrk~370 & 30.11.2003 & FPI501 & $120\times36$&   2.6     \\
    Mrk~600 & 01.12.2006 & FPI501 & $120\times36$&   2.2    \\
  \hline
\end{tabular}
\end{table*}

\subsection{Construction of maps  and measuring velocity dispersion}

We define the velocity dispersion of ionized gas $\sigma$ as the
standard deviation of the Gaussian profile fitted the H$\alpha$
emission line after accounting for the FPI instrumental profile and
subtracting the contribution of thermal broadening in the HII regions.
The procedure to measure $\sigma$ is described in detail in
\citet{MoiseevEgorov2008}.  In short, the observed profiles of the H$\alpha$ line were fitted by
the Voigt function, which is a convolution of Lorentzian and Gaussian
functions corresponding to the FPI instrumental profile and broadening of
observed emission lines respectively. The FWHM of instrumental profile
was estimated each night from Lorentzian  fitting of the He-Ne-Ar
calibration lamp emission scanned with FPI \citep{MoiseevEgorov2008}. 
The results of profile fitting were used to
construct two-dimensional line-of-sight velocity fields of ionized
gas, maps of line-of-sight velocity dispersion, free from the
instrumental profile ($\sigma_{obs}$) influence, and also the images
of galaxies in the H$\alpha$ emission line and in the continuum.

The accuracy of velocity dispersion was estimated from the
measurements of the S/N using the relations given in Figure~5 of
\citet{MoiseevEgorov2008}. On the $\sigma$ maps we masked out the
regions with a weak signal, where the formal error of velocity
dispersion measurements exceeded $10-12\km$ (which corresponds to
$S/N\leq5$). Emission line intensity maps were
constructed even for regions where the signal-to-noise ratio was
smaller: $S/N\approx2-3$.

The correction from the measured $\sigma_{i, obs}$ to the
  final $\sigma_i$, where $i$ corresponds to the pixel number,  was
done according to the relation \citep{Rozas2000}:
\begin{equation}
\sigma_i^2=\sigma_{i,obs}^2-\sigma_{N}^2-\sigma_{tr}^2, \label{eq1}
\end{equation}
where $\sigma_{N}\approx3\km$ and  $\sigma_{tr}\approx9.1\km$
correspond to the natural width of the emission line and its
thermal broadening at the temperature of $10^4$~K.

In most of the objects the emission line spectrum is very well
described by a single-component Voigt profile.  Only a few galaxies
have areas where the emission line profile has a complex (usually
two-component) structure, showing the presence of expanding shells
around the regions of star formation or possible supernova
remnants. In addition to the previously reported cases of UGC 8508,
UGCA~92 \citep{MoisLoz2012}, and SBS 0335-052E \citep{Moiseev2010},
compact regions with a two-component \Ha\ profile were found in UGC
260, UGC 1281, UGC 7047 and UGC 7651.

The mean velocity dispersion for the whole galaxy, weighted with
intensity, was calculated as:
\begin{equation}
\sigma= \frac{ \sum\sigma_i I_i}{\sum I_i}, \label{eq2}
\end{equation}
where  $I_i$ is the observed emission line  in the $i$-th pixel.

\begin{table*}
\caption{Integral parameters of galaxies}\label{tab_dat}
\begin{tabular}{lcccclcr}
\hline
    Name  &  D     &  $M_B$ &  $M_K$ &$\log\,L_{H\alpha}$& $\sigma$ &    $i$  &$V_{max}$ \\
          &  Mpc   &        &        &   $\mbox{erg}\,\mbox{s}^{-1}$ & $\km $   & deg.   & $\km$    \\
\hline
\multicolumn{8}{c}{Local volume subsample}\\
CGCG 269-049 &   4.59 & -13.11 & -15.46 &  37.24           & $13.6\pm 1.9$ &        43  &       9.8   \\
       DDO~53 &   3.56 & -13.37 & -15.00 &  38.93           & $21.0\pm 1.8$ &        27 &      16.4   \\
       DDO~68 &   9.80 & -15.27 & -17.15 &  39.33           & $19.9\pm 3.9$ &        65 &      57.2   \\
       DDO~99 &   2.64 & -13.52 & -15.26 &  38.44           & $19.2\pm 2.4$ &        52 &      11.7   \\
      DDO~125 &   2.74 & -14.33 & -16.97 &  38.28           & $16.3\pm 2.3$ &        63 &      17.8   \\
      DDO~190 &   2.80 & -14.19 & -16.52 &  38.44           & $18.5\pm 2.9$ &        60 &      24.7   \\
        IC~10 &   0.66 & -15.99 & -17.90 &  40.73           & $17.6\pm 0.7$ &        31 &      52.8$^*$\\
      IC~1613 &   0.73 & -14.54 & -16.90 &  38.43           & $25.9\pm 1.3$ &        22 &      26.7$^*$\\
       KK~149 &   8.90 & -14.85 & -17.20 &  38.58           & $19.0\pm 4.1$ &        58 &      26.2   \\
       KKH~12 &   3.00 & -13.03 & -15.89 &  38.65           & $17.9\pm 4.5$ &        90 &      19.5$^*$\\
       KKH~34 &   4.61 & -12.30 & -14.65 &  37.18           & $11.6\pm 6.6$ &        55 &      12.8$^*$\\
       KKR~56 &   5.90 & -14.39 & -16.74 &  38.27           & $17.8\pm 6.3$ &      --   &         --  \\
      UGC~231 &  12.82 & -18.38 & -19.98 &  39.75           & $18.0\pm 3.8$ &        90 &      92.8   \\
      UGC~891 &   9.38 & -15.90 &   0.00 &  38.94           & $15.4\pm 5.9$ &        65 &      60.0   \\
     UGC~1281 &   4.97 & -16.07 & -15.51 &  39.07           & $16.7\pm 5.0$ &        90 &      56.4   \\
     UGC~1501 &   4.97 & -16.52 & -18.22 &  39.52           & $16.6\pm 2.4$ &        75 &      47.5   \\
     UGC~1924 &   9.86 & -15.80 & -17.41 &  38.66           & $14.5\pm 5.4$ &        90 &      50.6   \\
     UGC~2455 &   7.80 & -18.14 & -20.00 &  40.71           & $18.3\pm 2.6$ &        51 &      47.9   \\
     UGC~3476 &   7.00 & -14.27 & -16.62 &  39.22           & $16.5\pm 2.3$ &        90 &      47.3$^*$\\
     UGC~3672 &  15.10 & -13.89 &   0.00 &  39.30           & $18.3\pm 3.6$ &        56 &      67.8   \\
     UGC~5221 &   3.56 & -17.09 & -20.27 &  40.02           & $18.1\pm 1.6$ &        61 &      58.5$^*$\\
     UGC~5423 &   8.71 & -15.62 & -17.74 &  39.20           & $22.0\pm 2.2$ &        56 &      24.8   \\
     UGC~5427 &   7.10 & -14.48 & -15.50 &  38.75           & $21.3\pm 4.8$ &        55 &      54.1   \\
     UGC~6456 &   4.34 & -14.03 & -15.72 &  39.23           & $17.9\pm 1.3$ &        66 &      15.0   \\
     UGC~7047 &   4.31 & -15.07 & -17.42 &  39.25           & $15.3\pm 1.5$ &        46 &      37.5   \\
     UGC~7611 &   9.59 & -17.73 & -20.86 &  40.26           & $23.1\pm 2.5$ &        77 &      51.5   \\
     UGC~7648 &   5.80 & -16.72 & -18.26 &  40.01           & $18.9\pm 1.8$ &        55 &      78.1$^*$\\
     UGC~7651 &   5.80 & -19.42 & -21.50 &  40.93           & $22.8\pm 1.8$ &        47 &     129.2   \\
     UGC~8313 &   9.20 & -15.22 & -17.94 &  39.58           & $21.9\pm 2.0$ &        77 &      45.0   \\
     UGC~8508 &   2.69 & -13.09 & -15.58 &  38.43           & $13.3\pm 2.4$ &        51 &      32.6   \\
     UGC~8638 &   4.27 & -13.74 & -16.63 &  38.66           & $16.1\pm 1.9$ &        49 &      18.2$^*$\\
    UGC~11425 &   3.60 & -14.32 & -15.57 &  38.49           & $14.4\pm 0.0$ &        35 &      37.1   \\
    UGC~11583 &   5.90 & -14.32 & -16.67 &  38.35           & $14.6\pm 4.5$ &        90 &      46.7   \\
    UGC~12713 &  12.20 & -15.95 & -16.83 &  39.45           & $18.6\pm 2.2$ &        72 &      44.9$^*$\\
      UGCA~92 &   3.01 & -15.59 & -16.56 &  39.44           & $16.5\pm 3.1$ &        56 &      39.7$^*$\\
     UGCA~292 &   3.62 & -11.79 & -13.56 &  38.44           & $11.8\pm 1.7$ &        45 &      21.9   \\
\multicolumn{8}{c}{XMD subsample}\\
 HS~0822+3542 &  13.50 & -12.90 &   --   &  39.24           & $19.4\pm 0.7$ &        31 &    12.3   \\
 HS~2236+1344 &  86.40 & -17.04 &   --   &  41.02           & $28.0\pm 0.9$ &        35 &    21.8   \\
SAO~0822+3545 &  13.50 & -13.26 &   --   &  37.96           & $17.9\pm 5.4$ &        63 &    12.4   \\
SBS~0335-052E &  53.80 & -16.87 & -18.44 &  41.02           & $30.6\pm 1.6$ &        37 &    28.2   \\
SBS~0335-052W &  53.80 & -14.68 & -16.12 &  39.70           & $20.2\pm 2.7$ &        37 &    12.4   \\
 SBS~1116+517 &  23.10 & -14.73 &   --   &  39.97           & $27.4\pm 1.5$ &        50 &    10.4   \\
 SBS~1159+545 &  52.20 & -14.65 &   --   &  40.24           & $22.9\pm 1.3$ &        66 &     9.1   \\
SDSS~J1044+03 &  53.80 & -16.19 &   --   &  40.69           & $28.7\pm 1.9$ &        51 &     9.0   \\
      UGC~772 &  16.30 & -14.88 &   --   &  39.33           & $22.0\pm 2.9$ &        40 &    38.5   \\
      UGC~993 &  40.30 & -17.72 &   --   &  40.55           & $22.1\pm 2.5$ &        69 &    48.5   \\
Anon~J0125+07 &  40.30 & -16.20 &   --   &  39.46           & $25.0\pm 5.0$ &        64 &    22.9   \\
\multicolumn{8}{c}{BCDG subsample}\\
      II~Zw~40 &   9.69 & -18.29 & -17.89 &  41.14           & $32.5\pm 1.2$ &        60 &    50.5   \\
      II~Zw~70 &  19.12 & -16.56 & -18.61 &  40.47           & $25.7\pm 1.2$ &        76 &    35.0   \\
    III~Zw~102 &  22.71 & -19.24 & -22.91 &  40.85           & $31.7\pm 2.6$ &        60 &   108.4   \\
    III~Zw~107 &  78.09 & -19.53 & -22.04 &  41.37           & $41.1\pm 1.4$ &        51 &    25.1   \\
         Mrk~5 &  13.96 & -15.47 & -18.04 &  39.55           & $18.0\pm 1.6$ &        48 &    33.5$^*$\\
        Mrk~33 &  22.30 & -18.28 & -21.31 &  40.98           & $37.7\pm 2.4$ &        47 &    32.6   \\
        Mrk~35 &  15.60 & -17.76 & -20.14 &  40.43           & $27.7\pm 1.2$ &        27 &    63.6   \\
        Mrk~36 &  10.43 & -14.71 & -16.14 &  39.91           & $23.8\pm 1.2$ &        47 &    18.8   \\
       Mrk~297 &  65.10 & -21.16 & -23.49 &  41.65           & $36.7\pm 1.7$ &        40 &   121.8   \\
       Mrk~324 &  22.43 & -16.44 & -18.95 &  39.68           & $28.8\pm 5.1$ &        38 &   108.0$^*$\\
       Mrk~370 &  10.85 & -16.83 & -19.51 &  40.12           & $24.9\pm 2.0$ &        45 &    53.5   \\
       Mrk~600 &  12.81 & -15.43 & -17.43 &  39.78           & $18.7\pm 1.6$ &        59 &   35.5$^*$\\
\hline
\multicolumn{8}{l}{$^*$:  $V_{max}$ data from HyperLEDA}\\
\end{tabular}
\end{table*}

Figure~\ref{fig1_1} presents the results of our observations with the
scanning FPI: the image in the \Ha{} line, the velocity field,
and the velocity dispersion, corrected for the thermal
broadening and natural width according eq.(\ref{eq1}). The velocity fields usually have more  points than the
velocity dispersion maps since, at the same $S/N$ level,the  velocity is measured
with a higher accuracy than the line width.

\begin{figure*}
\centerline{\includegraphics[width=\textwidth]{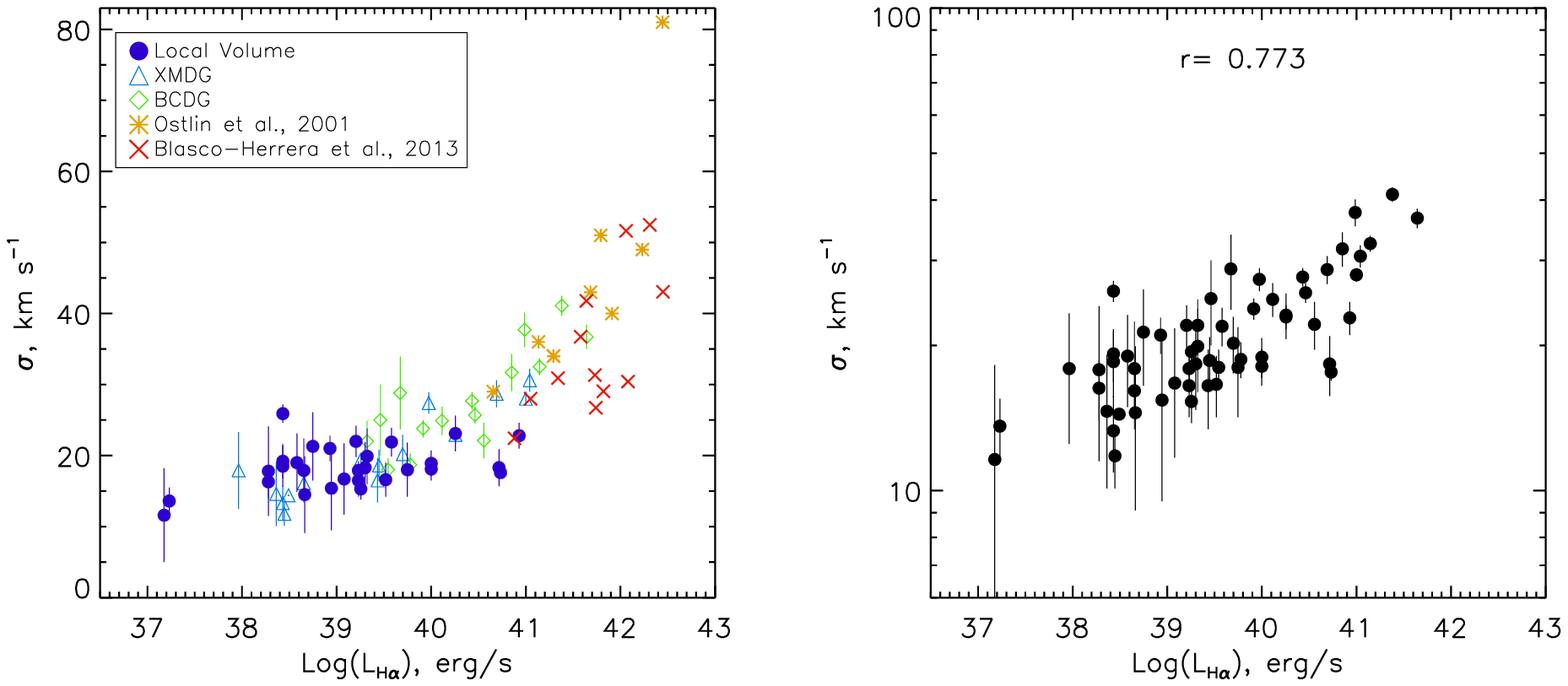}}
\centerline{\includegraphics[width=\textwidth]{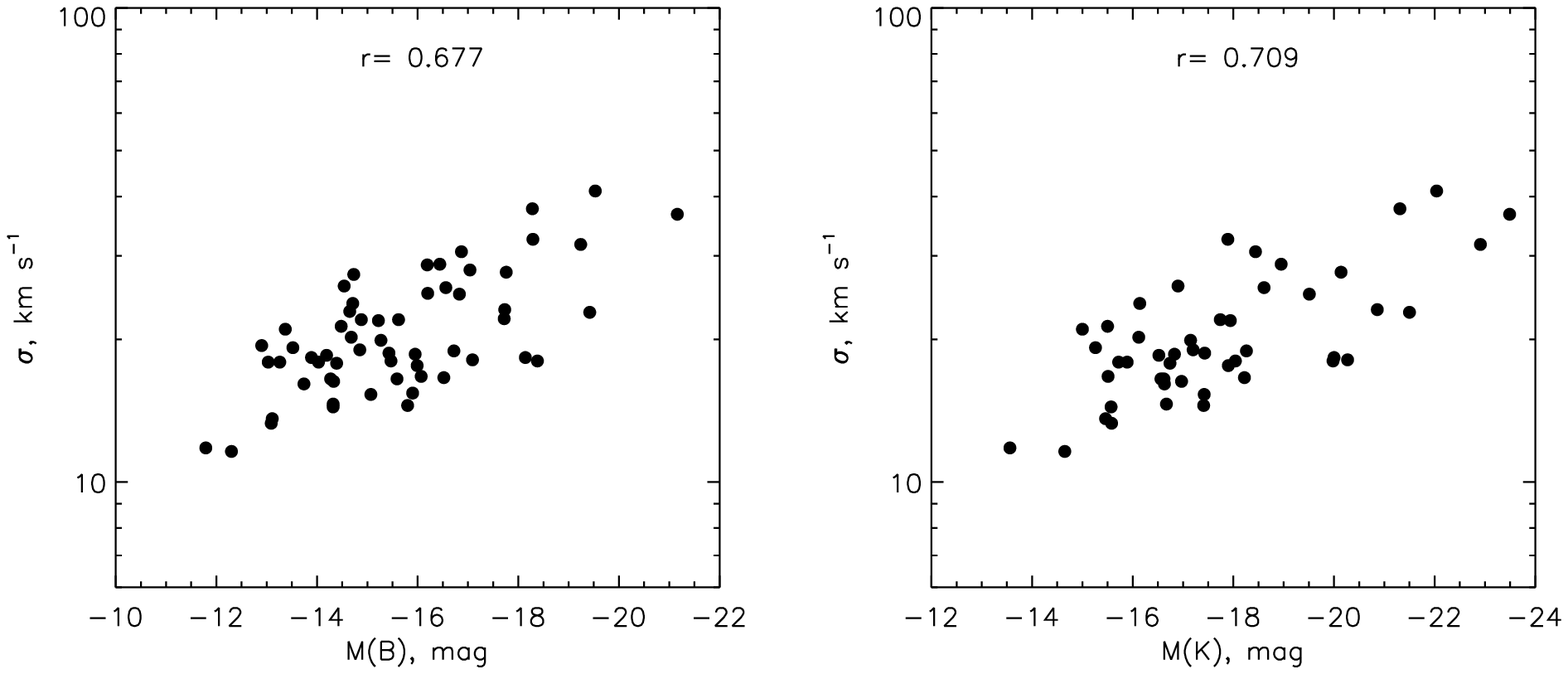}}
\caption{{\it Top}: The correlation of the average velocity dispersion
  $\sigma$ with \Ha{} luminosity. {\it Left}: Linear scale for
  $\sigma$. Color marks different subsamples and data from literature.
  {\it Right}: the same plot on the logarithmic vertical
  scale. Plotted are only the 6-m telescope data.  {\it Bottom}:
  Dependence of $\sigma$ on the absolute $B$ (left) and $K$ (right)
  magnitudes.  The correlation coefficient is marked in each
  panel. Error bars are not shown in the bottom panels to avoid
  crowding.  The velocity $\sigma$ correlates with the absolute
  magnitude, but the tightest relation is with the \Ha{} luminosity.}
\label{fig_lum1}
\end{figure*}

\begin{figure*}
\centerline{\includegraphics[width=\textwidth]{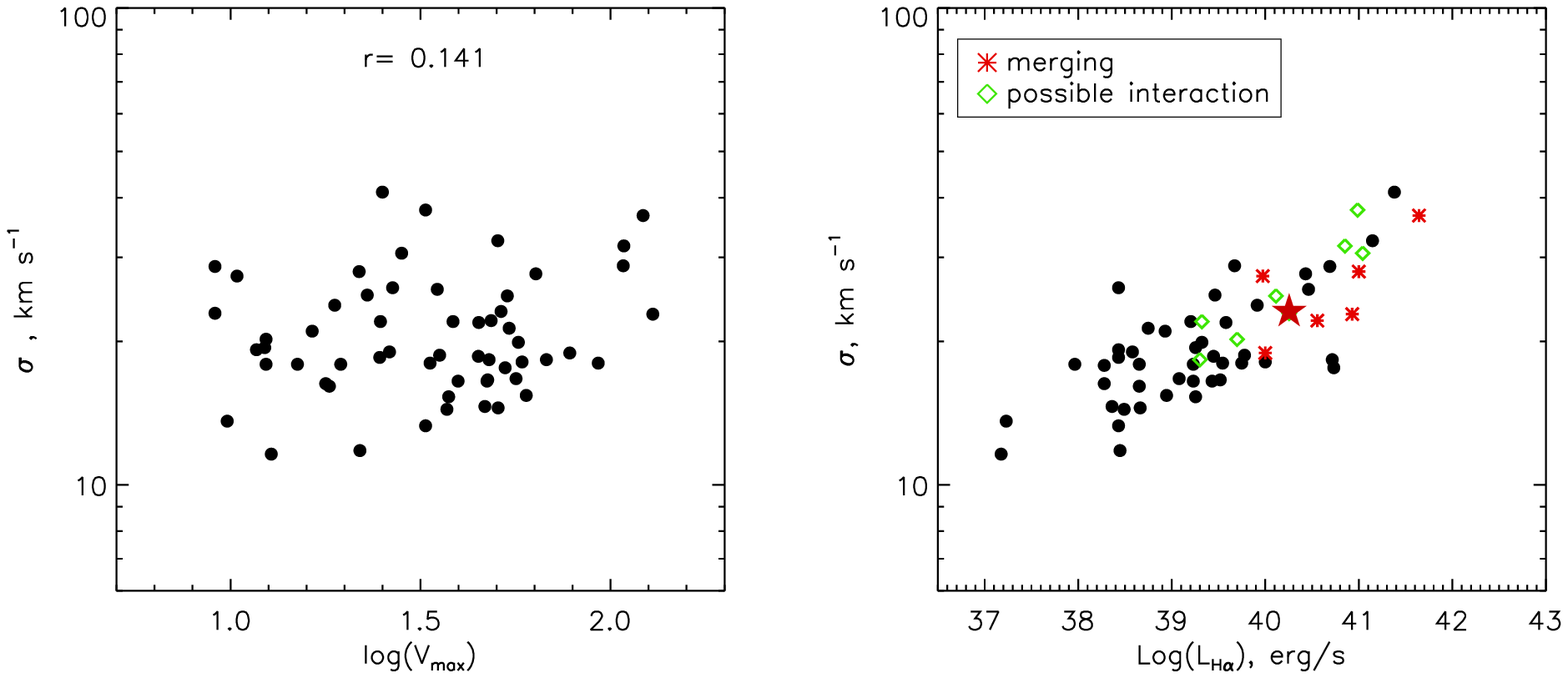}}
\caption{Left:  dependence of $\sigma$ on the amplitude of the
rotation curve. Right: the  $L_{H\alpha}-\sigma$ diagram, where
the colored dots mark the interacting galaxies. The asterisk
shows the  superwind galaxy UGC~7611 (NGC~4460).}
\label{fig_lum2}
\end{figure*}

\section{$L-\sigma$ relation}

\label{sec3}

The calculated kinematic parameters of the galaxies together with
their adopted distances, absolute magnitudes in the $B$ and
$K_s$-bands and total \Ha\, luminosity are given in the
Table~\ref{tab_dat}. We use the distances, luminosities in the $B$ and
$K_s$-bands, and in the \Ha\, emission line for the nearby galaxies
provided by the LVG database\footnote{http://lv.sao.ru/lvgdb/}
\citep{Kaisina2012}.  All luminosities are corrected for the internal
($A^i_B$) and Galactic ($A^g_B$) extinction, according to the values
given in this database\footnote{The absorption in the \Ha\, line was
 assumed to be $0.538(A^i_B+A^g_B)$, and in the $K_s$-band:
  $0.085(A^i_B+A^g_B)$.}

For nearby galaxies not listed in the LVG database (UGC~231, UGC~891,
UGC~1924, UGC~3672), the distances were adopted from
\citet{Karachentsev2004}.  For UGC~8313 we use data from
\citet{Tully1988}.  Apparent magnitudes of these galaxies were adopted
from RC3 \citep{RC3}; the $m_K$ values were retrieved from the 2MASS,
$A^i_B$ -- according to the relation of Verheijen (2001).  The flux in
the \Ha\, was taken from \citet{vanZee2000} and \citet{Kennicutt2008}. Distances and luminosities
for the XMD sample were taken -- as per \citet{Moiseev2010}, and for
the BCDG -- from \citet{Cairos2001}.

When estimating $L_{H\alpha}$, we take into account the fact that in
 observations with narrow-band filters  authors measure the
flux in \Ha+[NII]. The contribution of the nitrogen lines
[NII]$\lambda\lambda6548,6584$  was determined by an empirical
correlation, linking the [NII]/\Ha\, ratio with $M_B$
\citep{Kennicutt2008,Lee2009}. For the most of considered galaxies, fainter
than $M_B=-18$ this ratio is small ([NII]$/H\alpha<0.2$).

The inclination angles $i$ and maximal rotational velocities
$V_{\rm max}$ for the XMD sample are based on the results of
\citet{Moiseev2010}.  For most of Local Volume galaxies the data are
taken from \citet{Moiseev2014}. The kinematic parameters of III~Zw~102
and UGC~8313 were presented by \citet{Moiseev2008} and
\citet{Voigtlaender2015}.  For other galaxies $i$ and $V_{\rm max}$
were derived from the ionized gas velocity field in the same manner as
described in \citet{Moiseev2014}.  In cases, where \Ha\,rotation
curves never clearly come to a plateau, or velocity field of ionized
gas are dominated by non-circular motions (objects marked by asterisk in the Table~\ref{tab_dat}), we use estimates from the
HyperLeda \citep{Makarov2014}.

The velocity fields shown in Figure~\ref{fig1_1} reveal a component
related with regular rotating discs in a majority of the sample (61
per cents). The fraction of rotation dominated galaxies significantly
changes in the different subsamples. In the Local Volume subsample the
most of galaxies (72 per cents or 26 objects) have disc-like ionized
gas kinematics. Even among the remaining objects, in some dwarf
galaxies (DDO~53, UGC~6456, UGC~8638) appears a component
corresponding to a circular rotation. However, in these galaxiers
non-circular velocities have larger amplitude \citep{Moiseev2014}. In
these low-mass galaxies the \Ha-emitted gas is observed only in the
central region, where amplitude of their rotation curve is about
5--10$\km$. In contrast with the LV subsample, only half of XMD
galaxies (55 per cents or 6 objects) demonstrate disc-like rotation
gradient, including  cases of merger remains (HS~0822+3542 and UGC~772) which
reveal two rotating discs with different  orientation of spins.  This is not surprising because analysis of
\citet{Moiseev2010} provides the evidence for the crucial role of
interaction-induced star formation among galaxies in this subsample.  The similar
situation is also true  for BCDG subsample, where only 33 per cents of objects
(namely III~Zw~102, Mrk~33, Mrk~35 and Mrk~370) show disc-dominated
rotation in their velocity fields.

The maps of velocity dispersion $\sigma$ clearly demonstrate that in
the centre of star forming regions the velocity dispersion of ionized
gas has a minimum, whereas $\sigma$ increases towards periphery. Such a
feature in the distribution of the ionized gas velocity dispersion was
noted earlier in a number of studies
\citep[e.g.][]{Moiseev2010,Marino2013} and was discussed in
\citet{MoisLoz2012}, where such behavior is attributed to the
influence of young stellar groups on the surrounding gas.

\begin{figure}
\includegraphics[width=0.5\textwidth]{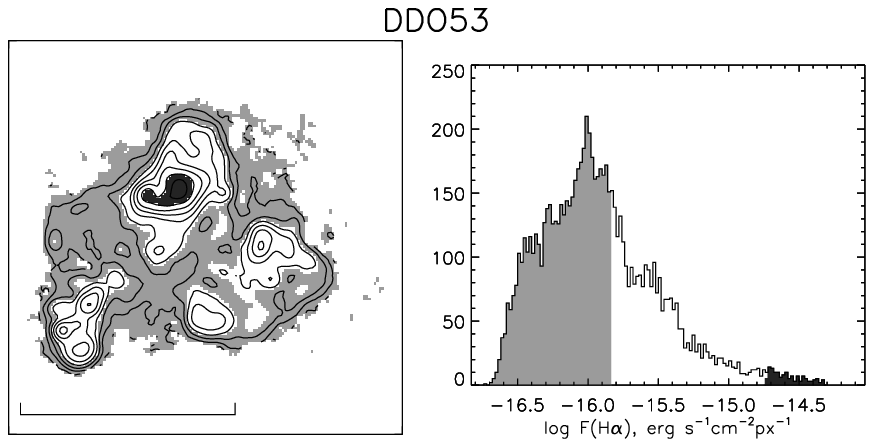}
\includegraphics[width=0.5\textwidth]{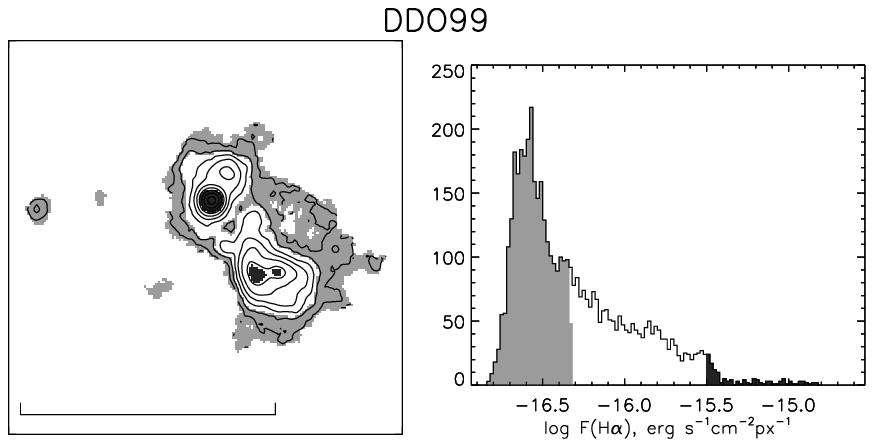}
\includegraphics[width=0.5\textwidth]{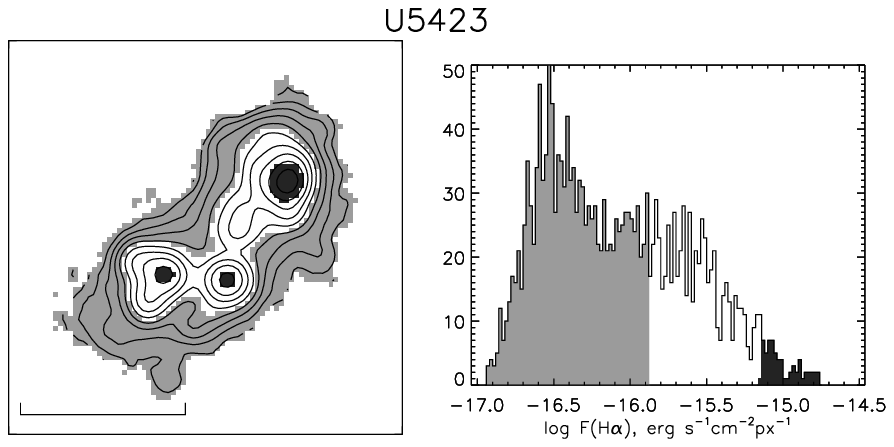}
\caption{Examples of \Ha~ ``cores'' (regions with very large
  $H\alpha$~fluxes, dark gray on the maps) and ``diffuse'' regions (low
  fluxes, light gray)  in DDO53, DDO99 and UGC 5423
  galaxies. The isophotes of the \Ha{} images are overlayed.  The
  horizontal bar shows the linear scale of 1 kpc.  To the right of
  each map we plot a histogram of the surface brightness distribution
  of the image pixels in \Ha. The ``core'' and ``diffuse'' components
  are filled with the same shades of gray.} \label{fig_diff}
\end{figure}

Figure~\ref{fig_lum1} shows different variants of $L-\sigma$
dependence for the galaxies in our sample. The literature most often
discusses the dependence of $\sigma$ on $L_{H\alpha}$ (or $L_{H\beta}$) , shown in the
upper panels. The top left figure separately shows all the three
subsamples of galaxies we have observed. It is clear that they form a
common sequence without any significant systematic offsets. Given
that, the galaxies from the LV in general have a lower luminosity than
the XMD and BCDG. For completeness, we present the same data for the
nine bright BCDGs from \citet{Ostlin1999, Ostlin2001}  and twelve
starburst galaxies from \citet{Blasco-Herrera2013} since the
measurement technique used by these authors is completely analogous to
ours -- averaging of the map of \Ha{} velocity dispersion, obtained
with a scanning FPI. Their
measurements complement our $L_{H\alpha}-\sigma$ sequence in the direction of
higher luminosities. Note that we have been able to significantly
continue the dependence on $L_{H\alpha}-\sigma$ towards the dwarf
galaxies, up to $L_{H\alpha}\propto10^{37}\,\mbox{erg}\,\mbox{s}^{-1}$,
while the vast majority of papers
\citep{TerlevichMelnick1981,BordaloTelles2011} considers the HII galaxies
with $L_{H\alpha}>10^{39}\,\mbox{erg}\,\mbox{s}^{-1}$ (with the
exception of papers devoted to various HII regions in the interiors of
large galaxies: \citet{Arsenault1988,Wisnioski2012}, and others). Not
to crowd the figures, further on we shall not show separately
subsamples nor depict the error bars.

The upper right panel shows the distribution of $L_{H\alpha}-\sigma$
in a more traditional form, on a logarithmic scale along both axes. It
can be clearly seen that the dependence between the logarithms of
luminosity and dispersion is almost exactly linear with a rather small
scatter.  This fact was noted in many studies, but, as already
mentioned in the Introduction,  in the case of giant HII regions it was usually  associated with
the virial ratio, i.e. by the fact that the ionized gas velocity
dispersion is definitely related with stellar velocity dispersion and is
controlled by the total mass of the system. However,  the luminosity in the Balmer emission lines  is
not a unique function of mass, while it is  determined by the
number of young OB-stars \citep[][and references therein]{Kennicutt1998}. While for any reasonable initial stellar
mass function, the total stellar mass is determined by the more
numerous but less massive stars that can not ionize the surrounding
gas.  Also the mass of the dark matter correlates exactly with this
stellar mass through the Tully-Fisher relation. If $\sigma$ is defined
by the mass of the system, it should better correlate not with
$L_{H\alpha}$, but with the luminosity in broader spectral bands, or
other parameters that are directly related to the mass.

But this is not observed. Figure~\ref{fig_lum1} shows the $L-\sigma$
dependence, built for $M_B$. In the optical $B$-band the older stellar
population has a larger  contribution, compared to the \Ha{}
luminosity. However, the point spread here is larger, and the
correlation coefficient is smaller: $r=0.68$ versus $r=0.77$ in the
case of of $L_{H\alpha}$. The $K$-band luminosity is directly related
with the total mass of stellar population of the galaxy, moreover, the
effect of interstellar reddening is much smaller here. However, the
point spread on the $M_K-\sigma$ diagram is same with $M_B$  ($r=0.71$, Figure~\ref{fig_lum1}
bottom right).

We  also  use a parameter  related to the dynamical mass of the system -- the amplitude of the rotation velocity,  $V_{\rm max}$.
Left panel in Figure~\ref{fig_lum2} indicates that ionized gas
velocity dispersion does not depend on the amplitude of the rotation
velocity $V_{\rm max}$ with the correlation coefficient being very low
($r=0.14$).  For galaxies with $V_{\rm max}=10-50\km$ the gas velocity
dispersion is remarkably large. On average $\sigma\approx 20\km$ with
substantial galaxy-to-galaxy variations. \citet{Green2010} came to a
similar conclusion that $\sigma$ is not related to the total stellar
mass for more distant galaxies with violent star formation.

\begin{figure*}
\centerline{\includegraphics[width=\textwidth]{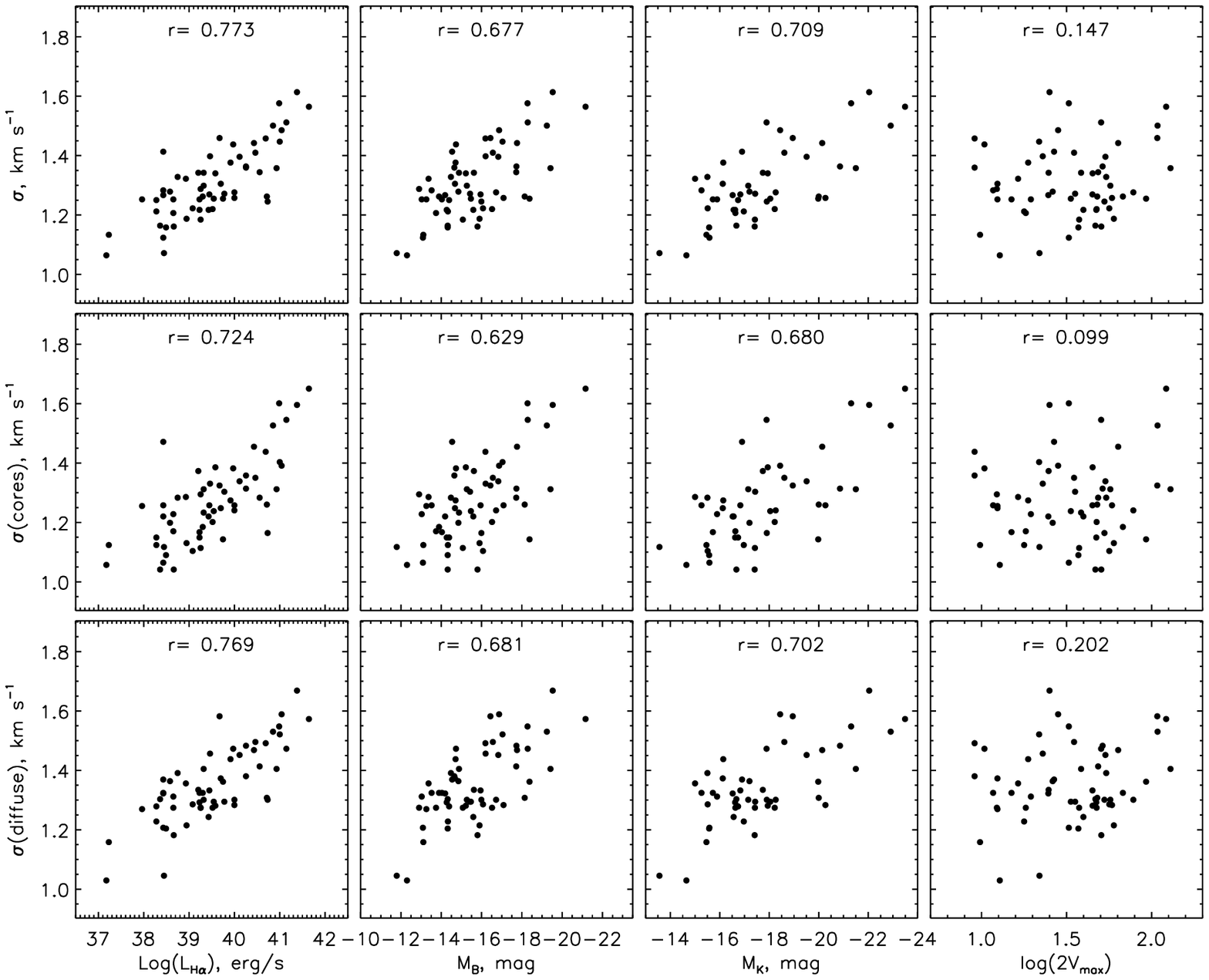}}
\caption{The $\sigma$ correlations with luminosity in different
ranges. Only the galaxies observed with the 6-m telescope. Top:
$\sigma$ was calculated from all pixels of the galaxy image.
Middle row: $\sigma$ was computed only for the cores of star
formation regions, where 20 per cents of the total  \Ha{} luminosity   are
concentrated. Bottom row: the same for the diffuse component
(20 per cents  of the total  \Ha{} luminosity in the low-brightness
``tail'' in the surface brightness distribution).} \label{fig_lum3}
\end{figure*}

Another argument against a direct relation of $\sigma$ with the galaxy
mass is that there should be observed a systematic departure from the
general trend for the galaxies in the transient unrelaxed state, such
as merger and interacting. We also attribute to clear mergers such
systems as Mrk~297 (two clearly kinematically decoupled components);
UGC~993,  HS~2236+1344 and SBS~1116+517 , where two rotating discs can be
identified in the velocity fields \citep{Moiseev2010}, and a tight
pair  UGC~7648/7651 (NGC~4485/4490) distorted by interaction. As `possible interaction' galaxies
we classify UGC~772, UGC~3672, SBS~0335-052W/E and SBS~1159+545  \citep[see the
arguments in ][]{Moiseev2010} and the galaxies with kinematically decoupled polar components: Mrk~33, Mrk~370
\citep{Moiseev2011} and IIIZw102 \citep{Moiseev2008}. However, in Figure~\ref{fig_lum2} all
these galaxies follow a general relationship, just like isolated galaxy UGC~7611 (NGC~4460), where the  ionized gas  is associated
with the circumnuclear  star burst and galactic wind outflow \citep{Moiseev2010_4460,Moiseev2014}.

Discussed above features of the $L-\sigma$ and $V_{\rm  max}-\sigma$ correlations suggest  that the relation between $\sigma$ and
$L_{H\alpha}$ (i.e. the current SFR) is primary. In this case  the correlations of velocity dispersion with
$M_B$, $M_L$, with the stellar and total mass are secondary, being the consequences of other scaling relations  in the
galaxies. Indeed, the more massive and bright  galaxies with  ongoing star formation as a rule
tend to have a greater luminosity in the Balmer lines as well.

\subsection{Line width $\sigma$ in dense  and diffuse gas}

\citet{Tenorio-Tagle1993} used an analytical model to support the idea
that $\sigma$ of ionized gas, observed in giant star-forming regions
and HII-galaxies is determined by the mass of these objects. The
authors conclude that $\sigma$ in the regions of greatest brightness
(``the kinematic cores of HII-regions'') is directly related to the
mass and size of the star forming region. At the same time, bright
HII-regions surrounded by the low-brightness coronae of ionized gas
with a larger velocity dispersion.  Such structure occurs as a result
of the influence of young stellar groups on the ISM.  But if the
velocity dispersion in the centres of star-forming regions determined
by the virial motions, then this value -- $\sigma({\rm core})$ should
be better correlated with the parameters related with mass ($M_B$,
$M_K$), rather than with $L_{H\alpha}$, which is controlled by the
number of ionizing photons from young massive stars. On the other
hand, the average velocity dispersion of diffuse environment should
show a clearer (than on the average for the entire galaxy) relation
with $L_{H\alpha}$, since not only the number of Lyman quanta, but
also the kinetic energy output of the winds of young stars and
supernovae is directly scaled with total number of OB stars \citep[see
Figure~1 in][]{Dopita2008}. While the gas velocity dispersion is
proportional to the square root of the kinetic energy of turbulent
motions.

We separate the velocity dispersion maps into cores of HII regions and
a diffuse component. For that we use a histogram of the brightness
distribution in the \Ha\, emission line for each galaxy.  We consider
as belonging to ``cores'' all the pixels, which contain the top 20 per
cents of the total \Ha{} luminosity, while the same 20 percent of the
luminosity in the low flux pixels are assigned to the diffuse
component.  Typical examples of galaxy \Ha\, map separation into two
components are shown in Figure~\ref{fig_diff} that also presents the
corresponding histograms of intensity of the surface brightness in
\Ha, explaining the method of separation (the bright and faint tail of
the distribution of surface brightness). We can see that ``cores''
really correspond to the very centres of bright HII regions, while
pixels marked as ``diffuse''  correspond to the envelopes
surrounding the regions of star formation.

For each of the components we calculate the flux-weighted velocity
dispersion. Figure~\ref{fig_lum3} shows the corresponding
relations. It is clear that the gas velocity dispersion of the cores
$\sigma({\rm core})$ and of the diffuse component
$\sigma({\rm diffuse})$ are very similar.  This is also confirmed by
the correlation coefficients with respective luminosities that are
almost the same, or even lower than those for the $\sigma$ of the
entire disc.

Therefore, the separation into the central and diffuse components
shows that the velocity dispersion at the centres of HII regions, as
well as the diffuse component are primarily determined by \Ha{}
luminosity, i.e. by the number of young massive stars. We have
performed this analysis for various ``core/diffuse'' separation
criteria. The general conclusion is the same as for the 20 percent
criterion described here.

\begin{figure*}
\centerline{\includegraphics[width=\textwidth]{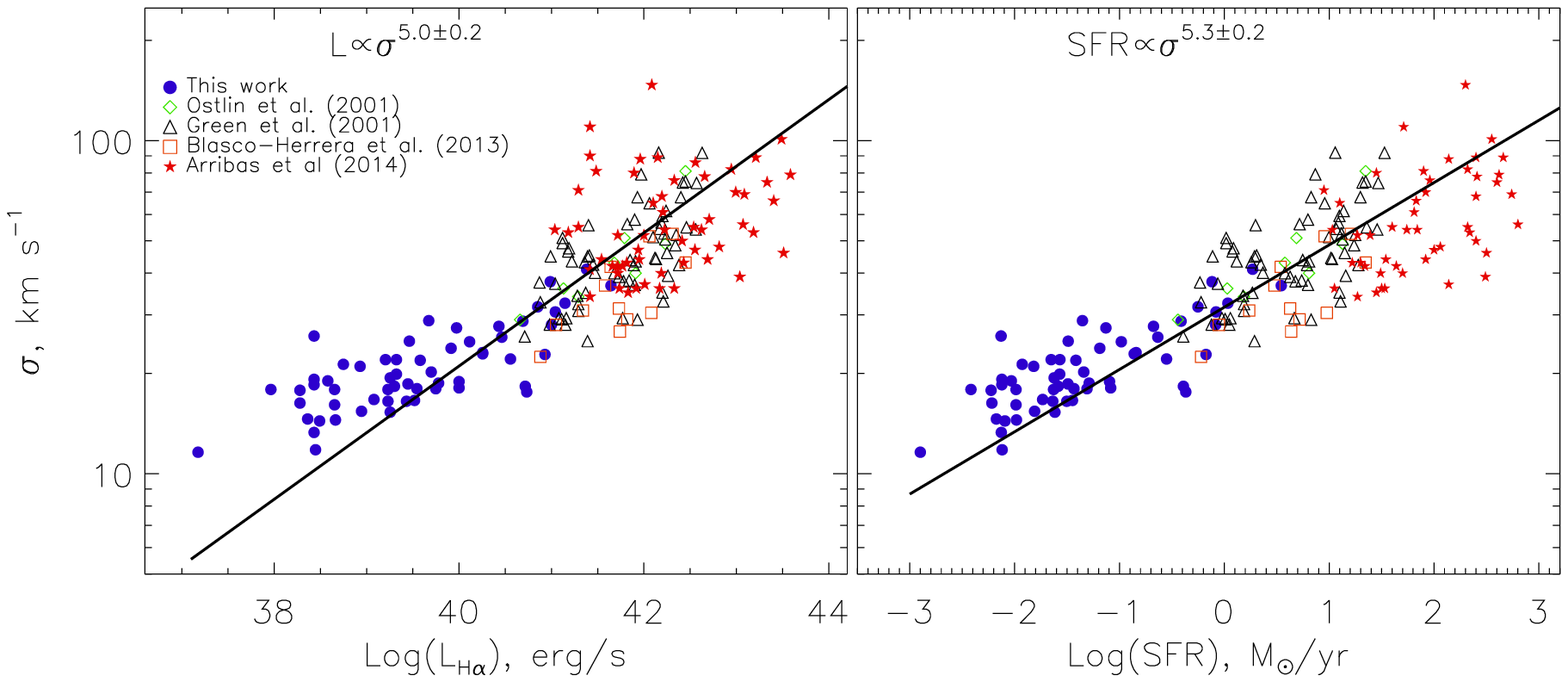}}
\caption{The $\sigma$ dependence on \Ha\, luminosity (left) and on
  $SFR$ (right) for different types of galaxies. Blue dots depict the
  galaxies of our sample, the other colors show the 3D-spectroscopic
  data of other authors for nearby starburst galaxies (see the
  legend). Black lines show a linear fit of these relationships in
  different regions.  
} \label{fig_lum4}
\end{figure*}

\section{Global $L-\sigma$ and SFR-$\sigma$ relations}

 As discussed in the Introduction, the $L-\sigma$ relation has
been extensively studied, mainly for intergalactic and extragalactic
HII regions and compact HII galaxies.  All the results agree that
there is a power-law relation: $L\propto\sigma^\alpha$. However, the
exponent $\alpha$ measured in different studies considerably varies
from $\alpha=3.15$ \citep{Roy1986} to 6.6  \citep{Hippelein1986}.
For discussion see also
\citet[][]{Blasco-Herrera2010, BordaloTelles2011,
  Wisnioski2012}. Recently \citet{Chavez2014}  obtained
$\alpha=4.65\pm0.14$ using a large sample of compact HII galaxies
restricted by velocity dispersion ($\log\sigma<1.8\km$) and equivalent
width ($EW(H_\beta)>50$\AA\, in order to minimize the contribution of
rotation-supported system and/or objects with a complex emission lines
structure. \citet{Swinbank2012} suggested $\alpha=3.8$ which provides
a common fit to the both local HII regions and starforming clumps in
high-redshift galaxies.  It is possible that different flux limits
used in observations to measure $\sigma$ for different types of
objects (e.g., individual HII regions, dwarf emission galaxies,
objects at large redshift) make some contribution to the uncertainty
with $\alpha$.  Potential biases of selecting bright $H\alpha$ regions
in the long-slit spectroscopy as compared with the full 3D
spectroscopy may also contribute to the disagreements. Indeed,
Figure~\ref{fig_lum3} shows that the velocity width
$\sigma({\rm core})$ in regions with high $H\alpha$ fluxes is
systematically smaller than $\sigma({\rm diffuse})$ in regions with
low fluxes. If one uses the spectra that are not deep enough, the
measured mean velocity dispersion will be underestimated. Another
possible reason for the disagreement is that the slope of the
$L(\sigma)$ relation changes with the luminosity and with the age of
the most recent burst of star formation episode.

 Figure~\ref{fig_lum4} combines our measurements with results by other
authors who used similar techniques to estimate the ionized gas
$\sigma$, i.e. 3D-spectroscopic observations (integral-field or
scanning FPI) and calculation of the flux-weighted velocity dispersion
for the whole galaxy instead individual HII regions: 9 blue compact
galaxies from \citep{Ostlin2001}, 11 starburst galaxies selected from
the SDSS for $z<0.03$ \citep{Blasco-Herrera2013}, 65 star-forming disc
galaxies at $z\sim0.1$ \citep{Green2010} and 57 local luminous and
ultra-luminous infrared galaxies (U/LIRGSs) without evidences of
active galactic nucleus \citep{Arribas2014}. The published data on
high-redshift systems were not included because the beam smearing of
the velocity gradient can produce a bias in the estimate of the
velocity dispersion \citep{Davies2011}.

The figure shows, that for a wide range of luminosities
$L_{H\alpha} = 10^{37}-10^{43.5} \,\mbox{erg}\,\mbox{s}^{-1}$, there is
a tight correlation between $L_{H\alpha}$ and $\sigma$. Using the least
square fit we derive the slope $\alpha=5.0\pm0.2$ for the present
$L-\sigma$ relation \footnote{Here  $\sigma$ is considered as an
independent variable in agreement with previous studies as it
discussed in \citet{BordaloTelles2011}.
The plots show that there is a
sort of kink at $\sim10^{41}\,\mbox{erg}\,\mbox{s}^{-1}$,
corresponding to $\sigma\sim30\km$.} This is why a linear
approximation of the $\log L_{H\alpha}-\log \sigma$ relation yields
substantially different slopes for large and small luminosities. 

Another way of presenting our results is to relate $L_{H\alpha}$
luminosity to the star formation rate SFR. It is known that the
$L_{H\alpha}$ luminosity is almost exactly proportional to the rate of
ongoing star formation (SFR) of young massive stars
\citep{Kennicutt1998}. However, a comparison of SFRs estimated
separately from $H\alpha$ data and \textit{GALEX} far-ultraviolet
observations shows that for the low luminosities this ratio is broken
due to the relative scarcity of massive stars in dwarf galaxies, in
relation with their initial mass function. \citet{Lee2009} show that
SFR$\propto L_{H\alpha}^{0.62}$ for
$L_{H\alpha}<2.5\times10^{39}\,\mbox{erg}\,\mbox{s}^{-1}$.

The right panel of Figure~\ref{fig_lum4} presents the SFR-$\sigma$
relation when we use \citet{Lee2009} conversion equations. The full
range of the SFR in considered objects is
$0.001-300\,M_\odot\,{\rm yr}^{-1}$.  For the U/LIRGSs sample we
accepted $SFR$ calculated from the near-infrared (IR) luminosity using
an equation from \citet{Kennicutt1998}, because a large uncertainties
with reddening correction of $L_{H\alpha}$ in these ``dusty'' systems
\citep{Arribas2014}. Indeed, the resulting scatter of points
corresponded to the high $\sigma$ for the $SFR$ seems to be smaller
than for the $H\alpha$ luminosities. Note that the combination of the
above factors (non-linearity of $SFR-L_{H\alpha}$ relation for the
faint dwarf galaxies, and IR-based $SFR$ for the most luminous ULIRGs)
leads a linear relation fitted as $SFR\propto\sigma^{5.3\pm0.2}$.

Using their data for the U/LIRGSs sample, \citet{Arribas2014} fitted
the similar relation as $\sigma\propto SFR^{0.12\pm0.03}$ that
corresponds to the slope $\alpha=8.3\pm2.1$. This value is significantly
larger than our measurements derived for the full sample included both
faint and ultra-luminous galaxies. \citet{Arribas2014} suggested that
relatively weak dependency of $\sigma$ on the total $SFR$, inferred from
their fit, is related with the fact that star formation is not a dominant
source driving the ionized gas velocity dispersion. Instead, they
present arguments in support the scenario where gravitational energy
associated with interaction and mergers has a significant contribution
to the gas turbulence for the $SFR >10\,M_\odot\,{\rm
  yr}^{-1}$.
However, our results clearly demonstrate the same tendency
$SFR-\sigma$ for the dwarf galaxies with and without interaction as
well as for U/LIRGs, which appear to have a larger fraction of ongoing
mergers. This fact may indicate that the role of interactions in driving
ionized gas turbulence in U/LIRGs was overestimated.

The other intriguing fact is that the slope of the $L-\sigma$ relation
inferred from our analysis ($\alpha\approx5.0-5.3$) is near the value
$\alpha=4.7$ obtained by \citet{Chavez2014}. From the one hand,
\citet{Chavez2014} considered individual giant HII regions and avoided
rotation-support system. From the other hand, the most of galaxies
presented on Figure~\ref{fig_lum4} reveal a domination of circular
rotation discs in their kinematics. In previous sections we presented
 arguments that the ongoing star formation is a dominant driver of
$\sigma$ calculated for the whole galactic disc. Why are the galaxies
of very different types and luminosities follow a similar $L-\sigma$
relation with the systems where ``the main mechanism of line broadening
is linked to the gravitational potential of the young massive cluster''
\citep{Chavez2014}?  Note that the range of luminosities on the
relation shown on the Fig.~\ref{fig_lum4} is twice larger as compared
with $L-\sigma$ considered in previous works for compact HII galaxies.

\section{Discussion}

\citet{Green2010} convincingly show that in a wide range of galaxy
luminosities, including the objects at $z=1-3$, the mean velocity
dispersion $\sigma$ is determined only by the star formation rate
(i.e. by the \Ha{} luminosity) and does not correlate with mass. In
this case, $\sigma$ is  characteristic of the energy injected in the
ISM by stellar winds, supernova explosions, and stellar radiation.
For the velocity dispersion of neutral gas, a similar conclusion --
weak correlation with galaxy mass -- was previously drawn by
\citet{Dib2006}. Further, the analysis of the shape of integrated HI
profiles by \citet{Stilp2013} shows that the velocity dispersion of a
broad component of the HI line in dwarf galaxies is defined by the
SFR$/M_{HI}$ ratio, although the dependence on the galaxy mass is also
present.

Previously we found that there is a close relationship between the
two-dimensional distributions of the line-of-sight velocity
dispersions of the ionized gas and the local \Ha{} luminosity
\citep{MoisLoz2012}. Specifically, we found that most of the areas with
the highest velocity dispersion belong to the diffuse low brightness
gas, surrounding the star forming regions. This contadicts the idea
that $\sigma$ is determined mainly by the distribution of mass in the
of galaxy. In this case one would have expected a different pattern --
the  maximum of velocity dispersion at the centres of star-forming regions and a decreasing $\sigma $
with the distance from the centre.

In the present paper we discuss different types of correlations of
$\sigma$ with integral parameters of galaxies -- the amplitude of the
rotation curve, and luminosities in different bands. We find that the
relationship with the parameters characterizing the mass of the galaxy
is considerably less distinct than the one with the ongoing star
formation, determined by the \Ha{} luminosity.   Moreover, the
  current SFR determines the magnitude of supersonic turbulent motions
  of gas not only in the starburst galaxies, but also in objects with
  a very low SFR up to $10^{-3}\,M_\odot\,{\rm yr}^{-1}$.

We analyse  two-dimensional velocity fields of ionized gas using
observations at the 6-m BTA telescope. This allows us to confidently
  measure the mean $\sigma$ across a galaxy, and study details of its
  distribution inside and outside of star-forming regions.  Our new
  data significantly extend the published $L-\sigma$ relations to the
  low mass galaxies  and provide the observational evidence
    that the star formation determines the velocity dispersion of the
    ionized gas:

\begin{itemize}

\item The ionized gas velocity dispersion, luminosity-averaged across
  the galaxy, is better correlated with the luminosity in the \Ha{}
  line than in the broad $B$ and $K$-bands.  There is almost no
  correlation of the velocity dispersion$\sigma$ with the rotation velocity of
  galaxy.

\item The gas velocity dispersion $\sigma$ in the cores of star-forming regions
       is nearly the same as $\sigma$ in the duffuse component with low \Ha\, fluxes.

\item There a common  $SFR-\sigma$ relation for the local
    galaxies in a very broad range of luminosities
    $L_{H\alpha} = 10^{37}-10^{43.5}$ that corresponds
    $SFR=0.001-300\,M_\odot\,{\rm yr}^{-1}$. The  fit of this
    relation $\sigma\propto SFR^\alpha$ provides the slope
    $\alpha=5.3\pm0.2$.

\end{itemize}

We therefore conclude that velocity of turbulent motions of ionized
gas in  galaxies is defined mainly by the energy that is
transferred to the interstellar medium from young stellar populations in
the form of ionizing radiation pressure, and by the winds of young
stars and supernova explosions. We believe that this conclusion is
important for both simulations of galaxy formation  and
for interpretation of the apparent emission line widths in galaxies
affected by various processes (e.g., star formation, merging, and
virial motions).

\section*{ACKNOWLEDGMENTS}
We are very grateful to the anonymous referee, and to Santiago Arribas, Eduardo Telles
and Roberto Terlevich for their constructive comments and suggestions
that helped us to improve and clarify our result.  We also thank Sean
Markert for discussions and comments. Our observations were done with
the 6-m telescope of the Special Astrophysical Observatory of the
Russian Academy of Sciences. We grateful to the staff of the
Observatory and specially Victor Afanasiev for his  great contribution to spectroscopy at the 6-m telescope. The observations were carried out with the financial
support of the Ministry of Education and Science of the Russian
Federation  (agreement No. 14.619.21.0004,   project ID  RFMEFI61914X0004).  We
have used the NASA/IPAC Extragalactic Database (NED) which is operated
by the Jet Propulsion Laboratory, California Institute of Technology,
under the contract with the National Aeronautics and Space
Administration. We acknowledge the usage of the HyperLeda database (http://leda.univ-lyon1.fr).
This work was supported by the Ministry of Education
and Science of the Russian Federation (project 8523) and by the
Research Program OFN-17 of the Division of Physics, Russian Academy of
Sciences. AM is also grateful for the financial support of the
non-profit ``Dynasty'' Foundation.  AK acknowledge support of NASA and
NSF grants to NMSU.

{}

\label{lastpage}

\end{document}